\documentstyle[11pt]{article}

\input{psfig}

\def\pa{\partial}
 
\def\g{\gamma} 
 
\def\b{\beta}

\def\l{\lambda} 
\def\m{\mu} 
\def\n{\nu}

\def\longrightarrow{\relbar\joinrel\relbar\joinrel\rightarrow}
\def\be{\begin{equation}}
\def\ee{\end{equation}}
     
\renewcommand{\baselinestretch}{2.0}

\begin{document}

\begin{flushright}
BRX TH-429\\\vspace{-.2in}
HUTP-98/A05
\end{flushright}

\vspace{-.4in}

\begin{center}
{\Large\bf The Non-Abelian Coulomb Phase of the \\
Gauged Vector Model at Large N}

\vspace{.15in}

\renewcommand{\baselinestretch}{1}
\small
\normalsize
Henric Rhedin\footnote{Supported by the Swedish Natural
Science Research Council (NFR) under grant no.
F-PD10883-305.}\\
Martin Fisher School of Physics\\
Brandeis University, Waltham, MA 02254\\
\vspace{.05in}

and\\

\vspace{.05in}
Howard J. Schnitzer\footnote{Research supported in 
part by the DOE under grant DE--FG02--92ER40706.}\\
Lyman Laboratory of Physics\\
Harvard University\\
Cambridge, MA 02138

and

Martin Fisher School of Physics\footnote{Permanent address.\\
\hspace*{.2in}rhedin, schnitzer@binah.cc.brandeis.edu}\\
Brandeis University, Waltham, MA 02254
\end{center}

\renewcommand{\baselinestretch}{2}
\small
\normalsize
\begin{quotation}
{\bf Abstract}: 
The renormalization group flows of the coupling constants for the gauged 
U(N) vector model, with $N_f$ massless fermions in the defining representation, 
are studied in the large $N$ limit, to all orders in the scalar coupling $\l$, 
leading order in 1/N, and lowest two orders in the gauge coupling $g^2$.  
It is shown that the restrictions of asymptotic freedom, and the reality of 
the coupling constants throughout the flows, places important restrictions 
on $N_f/N$.  For the case with massless mesons, these conditions are 
sufficiently restrictive to imply the existence of an infrared fixed-point 
$(g_*,\l_*)$ in both couplings.  Thus, the consistent massless theory is scale 
invariant, and in a non-abelian Coulomb phase.  The case of massive mesons, 
and of spontaneously broken symmetry is also discussed, with similar, 
but not identical, conclusions. Speculations related to the possibility that 
there is a non-perturbative (in $g^2$) breakdown of chiral symmetry are presented.
\end{quotation}

\noindent{\bf I. ~Introduction}

In spite of the enormous progress that has been made
in developing non-perturbative methods for supersymmetric
gauge theories \cite{001}, there remains considerable 
interest in approaches suitable for non-supersymmetric 
theories.  A frequently used technique for that purpose 
is the 1/N expansion for a theory with internal symmetry, 
such as SU(N) or O(N), say.  Examples include 't Hooft's 
analysis of gauge theories \cite{002}, string behavior in 
two-dimensional QCD \cite{003}, and O(N)-invariant 
$\l\phi^4$ theory \cite{004,005}.

The 1/N expansion for $\l\phi^4$ theory (in 3+1
dimensions) with O(N) or U(N) symmetry (the so-called
vector model) has been extensively studied as a 
renormalized field theory.  However, the renormalized
vector model encounters a number of problems \cite{005},
reviewed in ref. \cite{006}.

In \cite{006}, the U(N) vector model was extended by
gauging the theory, and adding $N_f$ massless
fermions in the defining representation.  The
model was considered to all orders in $\l$, and
leading order in 1/N and the gauge coupling $g^2$.
It was shown that this theory has two phases, one of
which is asymptotically free and the other not, with
the asymptotically-free phase consistent in that
it avoids the difficulties found by Abbott, 
{\it et al.} \cite{005}.  To the order considered in 
\cite{006}, the asymptotically free sector requires 
$0 < \l/g^2 < 4/3 \,
( N_f/N  -1)$ and $N_f/N < 11/2$ in the large $N$
limit.  If these conditions are not satisfied, one
returns to all the problems of the non-gauged model.

In this paper we examine the renormalization group
(RG) flows for $\l$ and $g^2$ for the same model, but
now extended to all orders in $\l$, leading order in
$1/N$, and lowest {\it two} orders in $g^2$.  Once
again one finds a range of parameters for which the
theory remains asymptotically free, and thus we 
concentrate on this phase, as this appears to be the
only consistent phase.  The contribution of the next
order in $g^2$ narrows the ``window"  for asymptotic
freedom somewhat.  We find that the scalar coupling
is bounded above by
\be
\l/g^2 \leq \frac{2}{3} \,
\left( \frac{N_f}{N} - 1 \right) +
\left[ \frac{4}{9} \, \left( 
\frac{N_f}{N} - 1 \right)^2 -3 \right]^{1/2}
\label{eq:1}
\ee
in the ultraviolet (UV) limit if the theory is to
be asymptotically free.  Since $\l /g^2$ in Eq. 
(\ref{eq:1}) must be real, this implies that
$(\frac{3\sqrt{3}}{2} +1) \leq (N_f/N)$.  It is also possible
to have an infrared (IR) fixed-point 
\cite{009} in $g^2$ if
\be
\frac{34}{13} < \frac{N_f}{N} < \frac{11}{2} \; ,
\label{eq:2}
\ee
in which case the gauge-coupling at the IR fixed 
point is
\be
\left( \frac{g_*}{4\pi} \right)^2 = 
\frac{~~~\left( \frac{11}{2} - \frac{N_f}{N}\right)}
     {13 \left( \frac{N_f}{N} - \frac{34}{13} \right)}
\label{eq:3}
\ee 
[Note that lower-bound of eq. (\ref{eq:2}) is less than
that imposed on $N_f/N$ by the reality of (\ref{eq:1}).]
Perturbation theory in $g^2$, can always be justified
if $N_f/N$ is sufficiently close to
11/2.  Further, for massless scalars, we explore a
possible infrared fixed-point for the scalar coupling
as well, which would mean that the theory is scale 
invariant at the infrared fixed-point $(g_*,\l_*)$.

Some consideration of cases with non-vanishing masses
are also presented.  This allows us to comment on the RG 
flows to the IR when various masses appear.  The results 
of this paper lead us to argue that the massless theory
is in a non-Abelian Coulomb phase, if parameters are
chosen such that the theory is asymptotically free.
An analogous discussion is also given for certain 
massive cases, as well as speculations related to a
possible non-perturbative (in $g^2$) breakdown of
chiral symmetry.

\noindent{\bf II. ~The Model}

Consider the theory of gauged complex scalar fields
in the defining representation of U(N), and $N_f$
massless fermions in the defining representation as
well, with Lagrangian \newpage
\begin{eqnarray}
N^{-1} {\cal L} & = & | \pa_\m \phi + ig
A_\m \phi |^2 + \frac{1}{2\l} \, \chi^2 \nonumber \\
& - &
\frac{\m^2}{\l} \, \chi - \chi |\phi |^2 
- \frac{1}{4} \, Tr (F_{\m\n} F^{\m\n} ) + 
i \sum^{N_f}_{i=1} (\bar{\psi}_i \g \cdot D \psi_i )
\label{eq:4}
\end{eqnarray}
In (\ref{eq:4}) $\phi$ and $\psi_i$ transform in the
defining representation of U(N), the gauge field
$A_\m$ in the adjoint, $\chi$ is a singlet, and 
$D$ is the covariant derivative.  [We do not write
gauge fixing terms explicitly.]  The field $\chi$
serves as a Lagrange multiplier, which if eliminated
reproduces the usual $\l\phi^4$ interaction.  The
coupling constants and fields have been rescaled so
that N is an overall factor of the Lagrangian,
and hence 1/N is a suitable expansion parameter.  
Note that there is no Yukawa coupling between $\phi$
and $\psi$, since both are in the defining
representation.

It is convenient to consider the model in Landau
gauge, so that the gauge parameter will not be
renormalized.  The renormalization of the theory
may be carried out in modified minimal subtraction.
This introduces an arbitrary mass-scale $M$ as a
result of the renormalization process.  [For more
details pertaining to the renormalization of this
model, see ref. \cite{006}.]  Particularly relevant
for us are the renormalized gauge and scalar
coupling constants $g(M)$ and $\l (M)$ respectively.
These coupling constants satisfy renormalization
group equations
$$
\b_g = M \, \frac{dg}{dM} \eqno{(5{\rm a})}
$$
and
$$
\b_\l = M \, \frac{d\l}{dM}  \eqno{(5{\rm b})}
\label{eq:5}
$$
It is possible to obtain these beta-functions to all
orders in $\l$, leading order in 1/N, and a perturbation
expansion in $g^2$, directly from the work of 
Machacek--Vaughn (MV) \cite{007}.  We evaluated this 
in the large N limit, and find \newpage
\renewcommand{\theequation}{\arabic{equation}}
\setcounter{equation}{5}
\begin{eqnarray}
16\pi^2 \b_g & = &
-g^3 \left( \frac{22}{3} - \frac{4}{3} \, 
\frac{N_f}{N} \right) \nonumber \\[.1in]
& - & \frac{4}{3} \, \frac{g^5}{(4\pi)^2} \,
\left( 34-13 \, \frac{N_f}{N} \right) + \ldots \; ,
\label{eq:6}
\end{eqnarray}
with $N_f/N$ fixed, and
\be
\b_\l = a_0 \, \l^2 - a_1\: g^2 \l + a_2\: g^4 \; ,
\label{eq:7}
\ee
where the $g^2$ dependent coefficients are 
\be
\left\{
\begin{array}{lcl}
(4\pi )^2 a_0 & = & 2 + 16 
\left( \frac{g}{4\pi} \right)^2 + \ldots \nonumber \\
(4\pi )^2 a_1 & = & 12 +  \frac{1}{3} \,
\left[ 256 - 40
\left( \frac{N_f}{N} \right)\right] 
\left( \frac{g}{4\pi} \right)^2  + \ldots \nonumber \\
(4\pi )^2 a_2 & = & 6 +  \frac{1}{3} \,
\left[ 304 - 64
\left( \frac{N_f}{N} \right)\right] 
\left( \frac{g}{4\pi} \right)^2  + \ldots
\end{array}
\right.
\label{eq:8}
\ee
Equations (\ref{eq:6})-(\ref{eq:8}) are
obtained by applying the MV results \cite{007} 
to our model.
Note that in (\ref{eq:6}) the scalar mesons make
no contribution to $\b_g$ in the large N limit, as
expected from 't Hooft's analysis 
\cite{002} of gauge theories
at large N.  Recall that conventional coupling 
constants have been rescaled as $g^2 N \rightarrow g^2$
and $\l N \rightarrow N$, (and fields rescaled as well)
so as to give the overall factor of $N^{-1}$ in
(\ref{eq:4}).  The remainder of the paper is devoted
to exploring the consequences of the coupled set of
coupled equations (5)--(\ref{eq:8}).

\noindent{\bf III. ~The Renormalization Group Flows}

Since $g(M )$ does not depend on $\l$, we may solve 
for it first.  Define
$$
t = ln \, M \eqno{(9{\rm a})}
$$
\vspace{-.1in}
and 
\vspace{-.3in}
$$
x(t) = g^2 (M) \; , \eqno{(9{\rm b})}
$$
then
\vspace{-.3in}
\renewcommand{\theequation}{\arabic{equation}}
\setcounter{equation}{9}
\be
\frac{dx}{dt} = -b_0 \, x^2 + b_1 \, x^3 + \ldots
\label{eq:10}
\ee
where
\vspace{-.3in}
\begin{eqnarray}
(4\pi )^2 b_0 & = & \frac{4}{3} \: 
\left( 11 - \frac{2N_f}{N}
\right) \nonumber \\
(4\pi )^2 b_1 & = & \frac{8}{3} 
\left( 13 \: \frac{N_f}{N} - 34 \right)
\label{eq:11}
\end{eqnarray}
If Eq. (\ref{eq:2}) is satisfied, then $g^2(M)$ is
asymptotically free, and there is an IR fixed-point
for $g^2$ given by Eq. (\ref{eq:3}).

In order to solve the equation $M \, \frac{d\l}{dM}
= \b_\l$, we first need to find an explicit solution
for $g^2 (M)$.  Make the change of variables
\be
ds = x(t) dt
\label{eq:12}
\ee
suggested by Calloway \cite{008}.  It is 
then straightforward
to show that 
\be
x(s) = 
\left[ A \: e^{sb_0} + \frac{b_1}{b_0}  \right]^{-1} \; ,
\label{eq:13}
\ee
where
\be
A = \left( \frac{1}{x_0} - \frac{b_1}{b_0} \right)
\label{eq:14}
\ee
with
\begin{eqnarray} 
x(s=0) \equiv  x_0  = g^2 (s=0)\; .
\label{eq:15}
\end{eqnarray}
The explicit relation between $t$ and $s$ is also
easily found to be
\be
t = \frac{A}{b_0} \, (e^{sb_0} - 1) + \frac{b_1}{b_0}
\, s \; ,
\label{eq:16}
\ee
so that $t$ versus $s$ is single-valued, and
$s \rightarrow \pm \infty$ when $t \rightarrow \pm
\infty$.  The integration constants were chosen so that
$t=0$ implies that $s=0$.  Then
$g^2_0 = g^2 (M_0)$, where $M_0$ is the mass-scale
at which coupling constants are defined.

Define 
\be
y(s) = \l (s)/g^2 (s)
\label{eq:17}
\ee
Then the renormalization group equation for $\l$
becomes, in terms of eq. (\ref{eq:13}),
\be
\frac{dy}{ds} = 
[a_0 y^2 - a_1 y + a_2 ] +
[b_0 - b_1 x (s) ] y
\label{eq:18}
\ee
In the ultraviolet (UV) limit $s\rightarrow \infty$,
$(g^2/4\pi ) \rightarrow 0$, so that for large $s$,
\be
\frac{dy}{ds} \simeq a_0\; y^2 - (a_1 -b_0)
y + a_2 \; ,
\label{eq:19}
\ee
where $b_0$ is given in (\ref{eq:11}), and 
$a_0, \; a_1,$ and $a_2$ are now just the leading
terms of (\ref{eq:8}).  It is convenient to write 
(\ref{eq:19}) as
\be
\frac{dy}{ds} = a_0
[y - y_+ (\infty )][y-y_- (\infty )]
\label{eq:20}
\ee
with
\be
y_\pm (\infty ) = \frac{2}{3} \,
\left( \frac{N_f}{N} - 1 \right) \pm
\left[ \frac{4}{9} \, \left( \frac{N_f}{N} \,
-1 \right)^2 - 3 \right]^{1/2} \; .
\label{eq:21}
\ee
Reality of the coupling constants in (\ref{eq:19})
requires
\be
(a_1 - b_0 )^2 \geq 4  a_2a_0 \; ,
\label{eq:22}
\ee
which is the requirement that (\ref{eq:21}) be real.
This implies, when combined with (\ref{eq:2}), that
\be
3.6 \simeq \left( \frac{3\sqrt{3}}{2}
 + 1 \right)  \leq N_f/N \leq \frac{11}{2} \; ,
\label{eq:23}
\ee
which therefore forces $b_1 > 0$.

Equation (\ref{eq:20}) implies the following
renormalization group flows to the UV, where
$y_0$ is an initial value with sufficiently large $s_0$
\footnote{More precisely in case (a) $y(s_0)$ lies on the boundary $y_+(s)$ 
defined in eq. (33).}.\\
a) ~If $y_0 = y_+ (+\infty )$, then $dy/ds = 0$ for
large $s$, and $y(s)$ remains at $y_+ (\infty )$ for 
increasing $s$.\\
b) ~If $y_0 < y_+ (+\infty )$, then $y(s) 
_{\stackrel{\longrightarrow}{\scriptstyle s
 \rightarrow \infty}} y_- (\infty )$.\\
c) ~If $y_0 > y_+ (+\infty )$, then $y(s) 
_{\stackrel{\longrightarrow}{\scriptstyle s 
\rightarrow \infty}} \infty $, and asymptotic 
freedom is lost.  Therefore
we see that $y_+ (\infty )$ defines the 
phase-boundary which
separates an asymptotic free theory (in 
{\it both} couplings) from the inconsistent 
non-asymptotically free theory. From (\ref{eq:21}), 
we see that this phase boundary is given by
\be
y_+ (\infty )  = 
\frac{2}{3} \left(\frac{N_f}{N} - 1 \right)
+
\left[ \frac{4}{9} \, \left(\frac{N_f}{N} 
-1 \right)^2 -3 \right]^{1/2} \; ,
\label{eq:24}
\ee
which then gives the upper-bound for $(\l /g^2)$
as stated in eq. (\ref{eq:1}).  [The inclusion of higher
order corrections in $g^2$ has reduced somewhat the 
upper-bound $\l /g^2 < 4/3 (N_f /N - 1)$ given in ref.
\cite{006}.]

We now turn to the consideration of the IR
region.  Let us set the renormalized value of the
mass-parameter $\m^2 /\l = 0$ in (\ref{eq:4}), so that
the scalar fields remain massless.  For this part
of the discussion, we assume that
the $N_f$ massless fermions do not acquire a mass by
means of a non-perturbative process.  [Later in the
paper we return to the possibility that
the fermions get a mass non-perturbatively.]  
With these assumptions, we can use 
the beta functions, as given by (\ref{eq:6}) and
(\ref{eq:7}), all the way to the IR limit,
$s \rightarrow -\infty$.

Since (\ref{eq:23}) requires $b_1 > 0$, we consider
the consequences of a fixed-point $g_*$. 
The IR limit of eq. (\ref{eq:18}) gives
approximately
\be
\frac{dy}{ds} \simeq a_0\, y^2 - a_1 \, y+a_2
\label{eq:28}
\ee
where the coefficients $a_0, \; a_1,$ and $a_2$ in
(\ref{eq:8}) are to be evaluated for $s \rightarrow
-\infty$.  [To obtain (\ref{eq:28}) note that
$x(-\infty ) = b_0/b_1$.]  The reality of
$(\l /g^2 )$ for $s\rightarrow -\infty$ requires
\be
a^2_1 - 4 \, a_2 a_0 \geq 0 \; .
\label{eq:29}
\ee
For convenience define $z = (g_*/4\pi )^2$.
Then using (\ref{eq:8}), (\ref{eq:29}) can be
cast into the inequality
\begin{eqnarray}
1 & + & \left[ 8.9 - 1.6 \left(\frac{N_f}{N}
\right)\right] z \nonumber \\[.1in]
& + & \left[ 8.3 - 9.5 
\left( \frac{N_f}{N} \right) + 1.9
\left( \frac{N_f}{N} \right)^2 \right] z^2
\geq 0 \; .
\label{eq:30}
\end{eqnarray}
Since we are considering the consequences of
a fixed point $g_*$, $N_f/N$ is not
independent of $z$.  One must use (\ref{eq:3})
to solve the constraints of (\ref{eq:30}).
Then (\ref{eq:30}) is satisfied for
all values allowed by (\ref{eq:23}), so that
\be
0 \leq z = \left(\frac{g_*}{4\pi}\right)^2
\leq 0.15 \; ,
\label{eq:31}
\ee
which justifies perturbation theory in $g^2$.
We have seen that for the massless theory,
$g^2 (s) _{\stackrel{\longrightarrow}{\scriptstyle s 
\rightarrow -\infty}} + \infty$ is {\it not} allowed
for our system of equations.  One {\it must}
have an IR fixed-point for $g^2$, to the order we are
working.  Higher correction in $g^2$ might alter
specific numerical values, but a change in 
qualitative conclusions would be surprising.

We also want to know whether $\l (-\infty )$ has a
fixed-point when one reaches the IR fixed-point
$g_*$.  Let us write (\ref{eq:28}) as
\be
\frac{dy}{ds} = a_0 [y - y_+(-\infty )]
[y-y_-(-\infty )] \; .
\label{eq:33}
\ee
Let $y (s_0)$ be an initial value for the solution
of (\ref{eq:33}), taken at sufficiently large negative $s_0$ 
\footnote{More precisely for cases (b) and (d) $y(s_0)$ lies on 
the boundaries $y_{\pm}(s)$ respectively, as defined in eq. (33).}\\
a) ~If $y(s_0) > y_+(-\infty )$, then 
$dy/ds > 0$, and $y(s)$ {\it decreases} towards
$y_+ (-\infty )$. [Of course that means $y(s)$
will increase as $s$ {\it increases}, which is
not asymptotically free, and thus
outside the consistent phase.]\\
b) ~If $y(s_0) = y_+ (-\infty )$, then 
$dy/ds = 0$, and $y$ remains at $y_+ (-\infty )$.\\
c) ~If $y_+ (-\infty) \geq y(s_0) \geq
y_- (-\infty), \; dy/ds < 0$, and $y(s)$
flows to $y_+ (- \infty )$ in the IR limit.\\
d) ~If $y(s_0) = y_- (-\infty )$, then $y$ remains at
$y_-(-\infty )$.\\
e) ~If $y(s_0) < y_-(-\infty )$, then $dy/ds > 0$,
and $y(s) _{\stackrel{\longrightarrow}{\scriptstyle s 
\rightarrow - \infty}}-\infty$, which is not allowed.\\  
Therefore, the values of $y_\pm (-\infty )$ are of
interest.  From (\ref{eq:33})
\be
y_\pm (-\infty ) = \frac{a_1}{2a_0} \pm \frac{1}{2a_0}
\, [a^2_1 - 4 a_0a_2 ]^{1/2}
\label{eq:34}
\ee
where the coefficients are evaluated in the IR
limit.  Since $y_\pm (-\infty ) > 0$, we see that
we reach a fixed point $\l_* = \l (- \infty )$ if
for very large negative $s_0$,
$y_- (-\infty ) \leq y(s_0 ) \leq y_+ (-\infty )$.
[Cases (b), (c) and (d).]  The actual values of
$(\l_*/g_*)$ are obtained from (\ref{eq:34}),
together with (\ref{eq:8}) evaluated at $g_*$
where the allowed range of $g_*$ is obtained from
(\ref{eq:31}).  Thus, consistency in the IR
of the massless theory, with no non-perturbative
generation of masses for the fermions,
leads to fixed points
$(g_*, \, \l_* )$ for the (gauge, scalar) 
couplings, and hence presumably to a scale-invariant
theory. At the IR fixed point $(g_*, \, \l_*)$ we have
massless interacting non-Abelian gauge bosons,
so that this is a non-Abelian Coulomb phase.
We shall argue below that this certainly occurs for
$N_f/N$ sufficiently close to 11/2.

Suppose now that the renormalized mass-parameter
in (\ref{eq:4}), $(\m^2/\l ) > 0$.  How does the
analysis of the IR region change?  In that case, all
the elementary scalar mesons acquire a mass.  Of
course, the criteria for the consistency of the 
theory in the UV does not change, but for momenta
$|p| << \m$, the mesons decouple, and the
effective low-energy theory is that of massless
gluons and fermions, again assuming that there is
no non-perturbation generation of masses for the 
fermions.  The scalar-coupling ``freezes"
at the renormalization-scale $M=\m$.  It is 
reasonable to take the arbitrary mass-scale 
$M_0 = \m$ as well.  With this convention, we have
\be
\l (s) = \l (s=0)\hspace{.2in} \mbox{\rm for} 
\hspace{.2in} s\leq 0 \; ,
\label{eq:35}
\ee
while the evolution of $g^2(s)$ towards the IR
is still governed by (\ref{eq:10})--(\ref{eq:16}).
In this case $g^2(s)$ will have an IR
fixed-point. Since consistency in the UV 
restricts one to the case where $N_f/N$ 
satisfies (\ref{eq:23}), $b_1 > 0$,
this is a non-Abelian
Coulomb phase as well.

Next suppose $(\m^2/\l ) < 0$ for the 
renormalized mass-parameter.  Then symmetry is
spontaneously broken to SU(N--1), with a Higgs 
boson which gives mass to some of the mesons and
gauge bosons.  The massless mesons and gauge bosons
transform as the defining and adjoint 
representations of SU(N).  Since there is no
Yukawa coupling, the $N_f$ fermions remain massless
in perturbation theory, and transform as the (N--1)
$\oplus$ 1 representation of SU(N--1).  The RG
flow to the UV is unchanged from our previous
analysis for momenta large compared to the 
masses.  In the IR, the RG flows are now appropriate
to SU(N--1).  These RG equations only differ from
those of the massless SU(N) theory by terms of 
order 1/N.  Thus, we expect an IR fixed-point
$(g_*,\l_*)$ for the massless SU(N-1) sector, as
the massive mesons and gauge bosons uncouple in
the low-energy region.

We have discussed the phase-boundaries in the far
UV and IR, which gave useful constraints on the
parameters of the model.  The phase-boundaries
can be described for arbitrary $s$, not just the
asymptotic values $\pm \infty$.  To do so,
consider (\ref{eq:18}) which can be written as
\be
\frac{dy}{ds} =
a_0 [y-y_+(s)][y-y_-(s)]
\label{eq:36}
\ee
where the $s$-dependent coefficients $a_0,\;
a_1$ and $a_2$ are given by (\ref{eq:8}),
and $b_0$ and $b_1$ by (\ref{eq:11}).  Therefore
\begin{eqnarray}
y_\pm (s) & = & 
\left[ \frac{a_1 -b_0 +b_1 \, x(s)}{2a_0}
\right] \nonumber\\
& \pm & \left\{\left[ \frac{a_1 -b_0 +b_1 \, x(s)}{2a_0}
\right]^2
- \frac{a_2}{a_0} \right\}^{1/2} > 0
\label{eq:37}
\end{eqnarray}
where reality of the couplings is required for all
values of $s$.  The curve $y_+(s)$ separates the
asymptotically free from the non-asymptotic free
phase.  Flows for $y(s) > y_+(s)$ grow in the UV
to $y(s) _{\stackrel{\longrightarrow}{\scriptstyle 
s \rightarrow \infty}} + \infty$, which is not
a consistent phase of the model.  Flows for
$y(s) < y_-(s)$ evolve in the IR to 
$y(s) _{\stackrel{\longrightarrow}{\scriptstyle s 
\rightarrow -\infty}}-\infty$, which is also not
permitted, as negative couplings are not allowed.
Therefore, $y_-(s) \leq y(s) \leq y_+ (s)$ is
required for consistency of the theory, which as
we have argued above describes a non-Abelian
Coulomb phase of the theory [see figure].

It should be mentioned, based on earlier 
work \cite{004}--\cite{006}, that there is a
singlet scalar meson bound-state in
meson-meson scattering.  This bound-state is
massive in the asymptotic free-phase, and
becomes massless on the phase-boundary
$y_+$, in accord
with the fact that $\l$ increases relative to 
$g^2$ as one approaches the phase-boundary $y_+$.
If $\l$ increases further (with $g^2$ fixed),
one leaves the asymptotic free phase, and the theory 
is inconsistent \cite{006} with all the problems of
the ungauged model \cite{005}.  In particular, the
bound-state becomes ``over-bound", {\it i.e.}, a
tachyon, if one is in the non-asymptotically free
sector.  Since there is no Yukawa coupling, this
scalar meson bound-state does not couple to the
fermions.

Finally we address the possibility that chiral
symmetry is broken non-perturbatively in $g^2$, and
that the fermions acquire a mass.  [It is not obvious
whether the scalars become massive by this mechanism,
as the scalars do not couple to the fermions.  We
shall discuss the consequences of the scalar becoming
or not becoming massive.]  Since the scalars do not
contribute to gluon dynamics in large $N$, we speculate
that the fate of the fermions is the same as that of
the theory without scalars.  Banks and Zaks \cite{009}
expressed the view that chiral symmetry is broken
spontaneously for all values of $N_f/N < 11/2$ consistent 
with asymptotic freedom.  However, more recently it
has been argued \cite{010,011} that there is a critical
value $(N_f/N)_{cr}$ above which the chiral symmetry
is restored.  Since we speculate that the value of 
$(N_f/N)_{cr}$ is not altered in large $N$
by the scalars of the model, as one would expect 
from the 't Hooft analysis \cite{002}
one obtains from eq. (94) of \cite{011} the 
estimate in the large $N$ limit
\be
( N_f/N )_{cr} = 4 \; .
\label{eq:38}
\ee
As the UV structure of the theory is not altered by
these considerations, we would have
\be
3.6 \simeq \left(
\frac{3\sqrt{3}}{2} +1 \right)
 \leq  N_f/N < 4 
\label{eq:39}
\ee
as the range of values
for which the fermions become massive, and
\be
4 \leq N_f/N < 11/2
\label{eq:40}
\ee
for which the fermions remain massless.  Let us
concentrate on the case $\m^2/\l = 0$.  Then
the scalars remain massless for $N_f/N$ in the
range (\ref{eq:40}), so that this is a massless
theory, with RG flow to the IR fixed-point
$(g_*,\l_*)$, as described by 
(\ref{eq:33})--(\ref{eq:34}),  a 
non-Abelian Coulomb phase.

Let us now consider the theory with $\m^2/\l =0$,
but with $N_f/N$ in the range (\ref{eq:39}); \\
(i)~ Assume that the scalars get a mass by the same
non-perturbative mechanism in $g^2$ by which the
fermions become massive.  Then, both the fermions
and scalars uncouple in the IR region, and the RG
flow towards the IR is just that of the pure
gluon theory.  This is presumably a confining 
phase.\\
(ii)~ The alternate possibility is that the scalars
remain massless in the phase described by (\ref{eq:39}).
If this is the case, only the fermions uncouple in
the infrared region.  For this possibility, the RG
flow for $s\rightarrow -\infty$ is described by
(\ref{eq:10}) and (\ref{eq:18}), but with
$(N_f/N) = 0$ in (\ref{eq:11}) and (\ref{eq:8}).  
Equation (\ref{eq:18}) evaluated in the IR region
now gives the reality condition,
\be
[a_1 + b_1 \: x(s)]^2 \geq 4 a_2 \: a_0 \; ,
\label{eq:41}
\ee
since $a_0, \; a_1, \; a_2$ increase as $x(s)$
as $s\rightarrow -\infty$. One may verify
that  (\ref{eq:41}) is {\it not} satisfied in the
IR! Thus, it appears that non-perturbative
generation of masses for the fermions, without a
concurrant generation of masses for the scalars,
is {\it not} a logical possibility.

In conclusion, we have studied the RG flows of the
gauged vector model, with $N_f$ massless fermions,
in the large $N$ limit, with particular emphasis on the
asymptotic free phase of the model, which is the only
consistent phase of the theory.  For the case of
massless mesons, we showed that an IR fixed point
$(g_*,\l_*)$ is extremely likely if certain required
restrictions on $N_f/N$ are met.  For the case of
massive mesons, the scalar meson sector decouples at
momenta small compared to the meson mass $\m$.  In
this case, the low-energy effective theory is that
of gauge bosons coupled to massless fermions.
Nevertheless, the restrictions on $N_f/N$ inherited
from the model in the UV region again make it likely
that the theory has a IR fixed-point $g_*$ for the
gauge coupling.  The case with spontaneously broken
symmetry to SU(N-1) in the large $N$ limit is closely
related to the massless SU(N) theory.  The predictions
of IR fixed points is predicated on the assumption
that higher order corrections in $g^2$ will not
qualitatively change our analysis.  This is certainly
true for $(N_f/N)$ sufficiently close to 11/2, where
 $g^2/4\pi$ is quite small. We also speculated on the
consequences of non-perturbative (in $g^2$) generation
of masses for the fermions.

One of us (HJS) wishes to thank the Physics Department
of Harvard University for its hospitality during the
spring semester of 1998.  He also wishes to thank
David Olmsted for his collaboration in the precursor 
to this work. Both authors wish to thank him  
for discussions. 

\newpage

\begin{figure} [h]
\centerline{
\psfig{figure=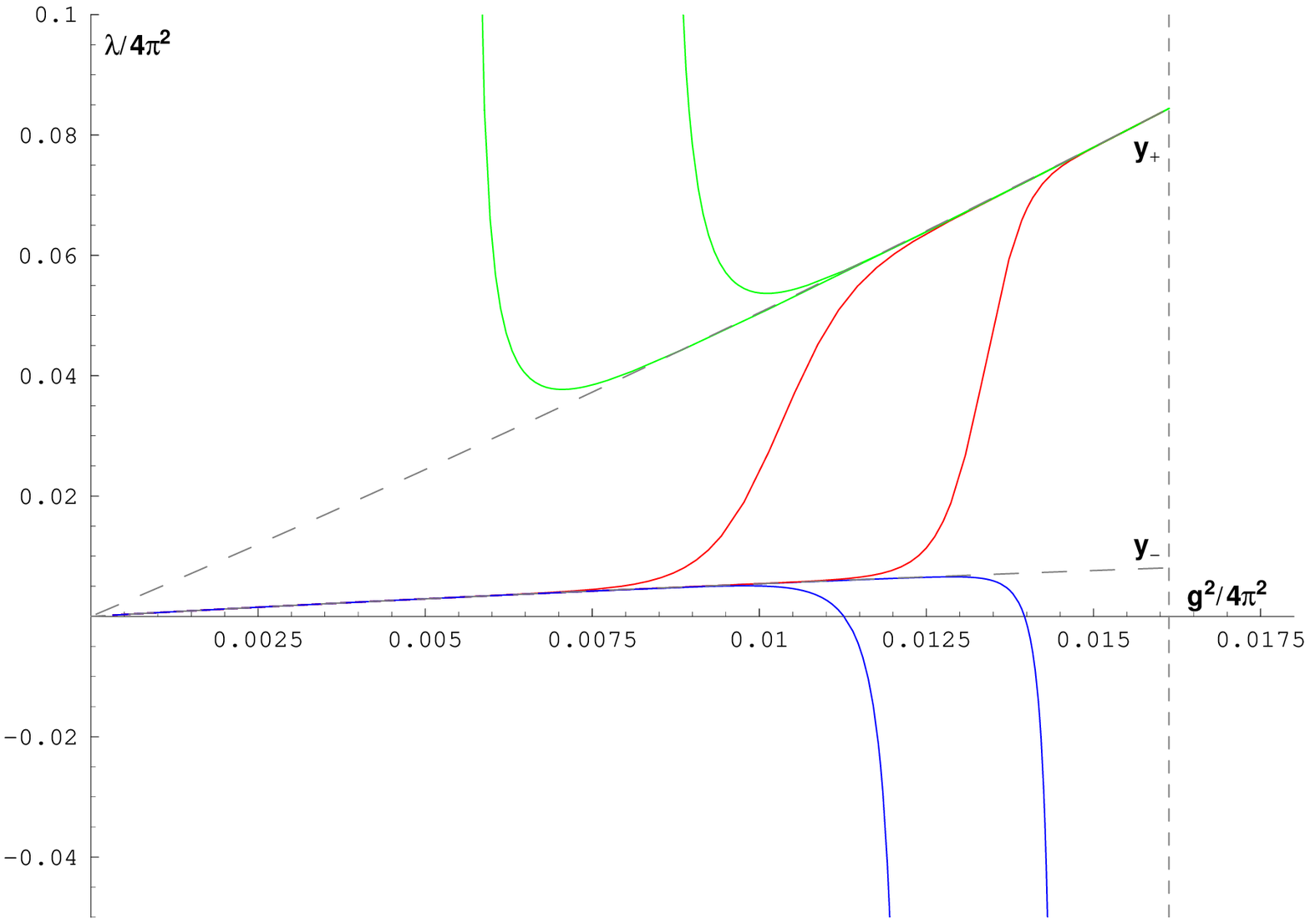,height=15cm,width=15.6cm}}
\end{figure}

Graph of the renormalization group flow for $N_f/N=5$, where 
ultraviolet to infrared flow progressing from left to right. 
The upper and lower dashed lines, $y_+$ and $y_-$ 
respectively, brackets flows consistent with asymptotic 
freedom and stability of the theory. The vertical dashed line 
at the right marks the value of the infrared fixed-point 
$g_{\star}$.

\end{document}